\documentclass[final,,twoside]{IEEEtranTCOM}

\normalsize

\ifCLASSINFOpdf

\else

\fi

\normalsize
\ifCLASSINFOpdf
\else
\fi

\usepackage{amsthm}
\usepackage{amsmath}
\usepackage{amssymb}
\usepackage{graphicx}

\usepackage{esint}
\usepackage{psfrag}
\usepackage{amsfonts}
\usepackage{cite}
\usepackage{balance}
\usepackage{caption}
\usepackage{subcaption}
\usepackage{epstopdf}
\usepackage{color}
\usepackage{afterpage}
\usepackage{tikz}
\usepackage{pgfplots} 
\usetikzlibrary{patterns}

\makeatletter

\theoremstyle{plain}

\makeatother

\providecommand{\theoremname}{Proposition}

\renewenvironment{IEEEbiography}[1]
  {\IEEEbiographynophoto{#1}}
  {\endIEEEbiographynophoto}

\begin{document}
%\markboth{\textit{IEEE Signal Processing Letters}}{Boulogeorgos, et al.:  Optimal Power Allocation for OFDMA Systems Under I/Q Imbalance}
%\markboth{\textit{IEEE Communications Magazine}}{Boulogeorgos, et al.:  Terahertz Technologies to Deliver Optical Network Quality of Experience in Wireless Systems Beyond~5G}
\title{Artificial Intelligence Empowered Multiple Access for Ultra Reliable and Low Latency THz Wireless Networks}

\author{Alexandros-Apostolos A. Boulogeorgos,~Senior Member, IEEE, Edwin Yaqub, Rachana Desai, \\
Tachporn Sanguanpuak, Nikos Katzouris, Fotis Lazarakis, Angeliki Alexiou,~Member, IEEE, and \\
Marco Di Renzo, Fellow, IEEE
%\\

\thanks{A.-A. A. Boulogeorgos and A. Alexiou are with the department of Digital Systems, University of Piraeus, Piraeus 18534, Greece. (E-mails: al.boulogeorgos@ieee.org, alexiou@unipi.gr).}
\thanks{E. Yaqub and R. Desai are with RapidMiner GmbH, Dortmund, Germany. (E-mails: \{eyaqub, rdesai\}@rapidminer.com).}
\thanks{T. Sanguanpuak is with Nokia Standardization Department, Oulu, Finland. (Email : tachporn.sanguanpuak@nokia.com).}
\thanks{N. Katzouris, and F. Lazarakis  are with the Institute of Informatics and Telecommunications, National Centre for Scientific Research–“Demokritos,” Athens, Greece. (E-mails: \{nkatz, flaz\}@iit.demokritos.gr).}
\thanks{M. Di Renzo is with Universit\'e Paris-Saclay, CNRS,
CentraleSup\'elec, Laboratoire des Signaux et Syst\`emes, 3 Rue
Joliot-Curie, 91192 Gif-sur-Yvette, France.
(marco.di-renzo@universite-paris-saclay.fr).}
\thanks{This work has received funding from the European Commission's Horizon 2020 research and innovation programme under grant agreements No. 871464 (ARIADNE), as well as the US National Science Foundation (NSF) under Grant No. CNS-2011411.}
}

\maketitle	

%\vspace{-2cm}
%\newpage
\begin{abstract}
Terahertz (THz) wireless networks are expected to catalyze the beyond fifth generation (B5G) era. However, due to the directional nature and the line-of-sight demand of THz links, as well as the ultra-dense deployment of THz networks, a number of challenges that the medium access control (MAC) layer needs to face are created. In more detail, the need of rethinking user association and resource allocation strategies by incorporating artificial intelligence (AI) capable of providing ``real-time'' solutions in complex and frequently changing environments becomes evident. Moreover, to satisfy the ultra-reliability and low-latency demands of several B5G applications, novel mobility management approaches are required. Motivated by this, this article presents a holistic MAC layer approach that enables intelligent user association and resource allocation, as well as flexible and adaptive mobility management, while maximizing systems' reliability through blockage minimization. In more detail, a fast and centralized joint user association, radio resource allocation, and blockage avoidance by means of a novel metaheuristic-machine learning framework is documented, that maximizes the THz networks performance, while minimizing the association latency by approximately three orders of magnitude. To support, within the access point (AP) coverage area, mobility management and blockage avoidance, a deep reinforcement learning (DRL) approach for beam-selection is discussed. Finally, to support user mobility between coverage areas of neighbor APs, a proactive hand-over mechanism based on AI-assisted fast channel prediction is~reported.  
\end{abstract}

%\newpage

\section{Introduction}\label{S:Intro}

Bringing the fiber quality of experience (QoE) into the wireless world is a pillar promise of beyond fifth generation (B5G) networks~\cite{Boulogeorgos2018a}. In this direction, the commercialization of terahertz (THz) wireless networks capable of supporting bandwidth-hungry applications, like holographic reality, collaborative robots, and self-driving vehicles, without compromising reliability or increasing latency, is a key objective~\cite{Petrov2018}. However, THz wireless systems are susceptible to line-of-sight (LoS) obstructions~\cite{Boulogeorgos2021b}. Moreover, due to the shorter ranges of THz frequencies, THz wireless networks are expected to be ultra-dense~\cite{Shafie2021,Zhang2021}. Thus, the congestion of static or dynamic objects/users would lead to LoS~blockages. 

This poses a number of important challenges, including user equipment (UE) to access point (AP) association, radio resource (RR) allocation, and mobility management (MM) that the THz wireless systems medium access control (MAC) need to address. In particular, a joint UE-AP association and RR allocation strategy capable of minimizing the LoS blockage, while maximizing the network performance, under an ultra-low latency demand is required. This inspired the design of joint UE-AP association and RR allocation strategies, via the implementation of metaheurestic~\cite{Boulogeorgos2018} or machine learning (ML) approaches~\cite{Feng2018,Wang2021}. However, the aforementioned contributions overlook an important factor, i.e., the negative impact of~blockage~\cite{Boulogeorgos2021}. 

From the MM point of view, two MAC-related research directions, namely beam selection and pro-active hand-over, have been followed~\cite{Boulogeorgos2021a}. In more detail, in~\cite{Ma2019}, a random forest (RF) classification-based beam-selection scheme was reported that alleviates the computational intensity of conventional approaches, while boosting performance in terms of data-rate maximization. In~\cite{Abuzainab2021}, a deep learning (DL)-based, proactive hand-over and beam-selection approach for THz drone communications was presented. Both the aforementioned approaches, require a  sufficient number of training samples and lack the ability to adapt to continuously changing wireless~environments. 

A MAC protocol for ultra-reliable and low-latency (URLL) THz wireless networks should, by-design, countermeasure against the effects of blockage, provide fast adaptation to changing environments, and support mobile UE (MUE). However, no holistic approaches that jointly address these challenges have been proposed. Motivated by this, this article aims to set the stage for the development of an intelligent MAC protocol that enables centralized UE-AP association, local (i.e., w.r.t. individual APs) and adaptable beam-selection, as well as pro-active hand-over. Specifically, the contribution of this work is as~follows:
\begin{itemize}
    \item We present a joint UE-AP association and RR allocation with blockage avoidance approach, based on a novel, hybrid metaheuristic-ML framework, capable of reducing the association latency by more than $3$ orders of~magnitude. 
    \item To support mobility management and blockage avoidance within the AP coverage area, we propose a DRL approach that enables flexible and fast~adaptation.
    \item Finally, to reliably support MUE that moves between AP coverage areas, we present a pro-active hand-over approach, built upon a novel AI-assisted fast channel prediction framework. 
\end{itemize}

\section{AI-assisted joint user association and RR allocation for blockage~avoidance}

\underline{Problem description}: The association and blockage minimization problem aims at providing URLL connectivity by quickly associating UEs to APs, while satisfying a number of hard and soft constraints. The difference between hard and soft constraints is that the first ones determine the feasibility of the solution, while the latter the quality of the solution. Three hard constraints have been considered, namely: i) unique assignment, ii) grouped allocation, and iii) RR allocation efficiency maximization. The first constraint ensures that each UE is associated to a single AP. Note that the association process requires that the AP has enough available RR to cover the UE data-rate demands. The second constraint enables the association of multiple UEs to a single AP. The final hard constraint aims to maximize the RR usage at APs, without exceeding their resource availability, while using the minimal number of APs. On the other hand, one soft constraint need to be satisfied, namely LoS maintenance. This is translated to LoS blockages avoidance or~minimization.

\underline{Why AI is needed}: Efficiently solving the aforementioned problem in dense and changing environments is expected to play a vital role in optimal RR management towards delivering high QoE. However, conventional approaches based on constraint solving and combinatorial optimization techniques cannot meet the latency requirements for connection establishment, given the scale, density, coverage, and RR management requirements. For instance, the association strategy in THz wireless systems requires LoS blockage avoidance/minimization. Even in networks of moderate size, this problem quickly becomes intractable for classical constraint solving techniques, due to the combinatorial explosion in the space of possible assignments. An alternative is to use ML-based techniques to train models that learn to approximately solve such optimization problems. Such techniques trade optimality for efficiency, since the solutions are typically sub-optimal. On the other hand, once trained, such a model may predict solutions on unseen instances of the optimization problem in a tiny fraction of the time that a constraint solver would need, thus fostering efficiency in tasks that require real-time responsiveness and adaptivity to change, such as the tasks that we address here.    

\underline{A hybrid metaheuristic-ML framework}:
We propose such an AI/ML-based approach to solve the UE-AP association problem, that combines classical search with ML. In particular, after formalizing the problem as a constraint satisfaction one and given a number of different instances thereof (snapshots), classical constraint solving techniques, assisted by metaheuristics to select among a rich library of sophisticated search algorithms, are used to derive optimal solutions to these snapshots. These solutions are then used to train and validate a classifier that learns to mimic the solver's behaviour by generalizing from the training snapshots, towards approximating solutions to unseen ones. As indicated above, the approach is generic enough to be applicable to any type of optimization problem, while, as we detail below, the it is capable of adapting to change, by retraining the model, on-the-fly, on new snapshots it exhibits low predictive performance on.      

\begin{figure*}
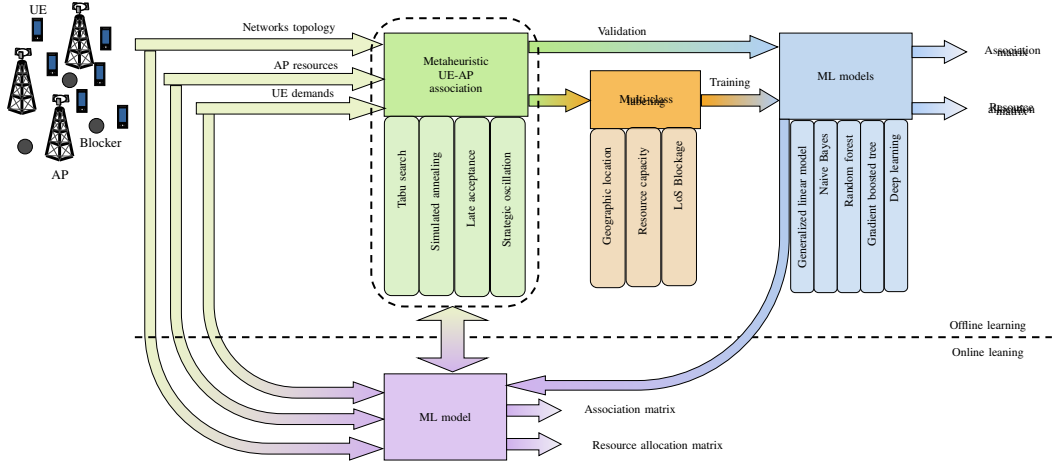

	\centering
	\scalebox{0.5}{
		
		% Gradient Info
		
		\tikzset {_91mg6cj4p/.code = {\pgfsetadditionalshadetransform{ \pgftransformshift{\pgfpoint{0 bp } { 0 bp }  }  \pgftransformrotate{0 }  \pgftransformscale{2 }  }}}
		\pgfdeclarehorizontalshading{_k3yyu2wlx}{150bp}{rgb(0bp)=(0.92,0.95,0.79);
			rgb(37.5bp)=(0.92,0.95,0.79);
			rgb(62.5bp)=(0.82,0.7,0.93);
			rgb(100bp)=(0.82,0.7,0.93)}
		
		% Gradient Info
		
		\tikzset {_dyp3rd50t/.code = {\pgfsetadditionalshadetransform{ \pgftransformshift{\pgfpoint{0 bp } { 0 bp }  }  \pgftransformrotate{0 }  \pgftransformscale{2 }  }}}
		\pgfdeclarehorizontalshading{_78u7f44gg}{150bp}{rgb(0bp)=(0.92,0.95,0.79);
			rgb(37.5bp)=(0.92,0.95,0.79);
			rgb(62.5bp)=(0.82,0.7,0.93);
			rgb(100bp)=(0.82,0.7,0.93)}
		
		% Gradient Info
		
		\tikzset {_gurhkrq59/.code = {\pgfsetadditionalshadetransform{ \pgftransformshift{\pgfpoint{0 bp } { 0 bp }  }  \pgftransformrotate{0 }  \pgftransformscale{2 }  }}}
		\pgfdeclarehorizontalshading{_sgz01oxso}{150bp}{rgb(0bp)=(0.92,0.95,0.79);
			rgb(37.5bp)=(0.92,0.95,0.79);
			rgb(62.5bp)=(0.82,0.7,0.93);
			rgb(100bp)=(0.82,0.7,0.93)}
		
		% Gradient Info
		
		\tikzset {_hp2xi0i66/.code = {\pgfsetadditionalshadetransform{ \pgftransformshift{\pgfpoint{0 bp } { 0 bp }  }  \pgftransformrotate{0 }  \pgftransformscale{2 }  }}}
		\pgfdeclarehorizontalshading{_rba7ft9e0}{150bp}{rgb(0bp)=(0.72,0.91,0.53);
			rgb(37.5bp)=(0.72,0.91,0.53);
			rgb(62.5bp)=(0.96,0.65,0.14);
			rgb(100bp)=(0.96,0.65,0.14)}
		
		% Gradient Info
		
		\tikzset {_5ts1nv8ue/.code = {\pgfsetadditionalshadetransform{ \pgftransformshift{\pgfpoint{0 bp } { 0 bp }  }  \pgftransformrotate{0 }  \pgftransformscale{2 }  }}}
		\pgfdeclarehorizontalshading{_4ukog80xo}{150bp}{rgb(0bp)=(0.72,0.91,0.53);
			rgb(37.5bp)=(0.72,0.91,0.53);
			rgb(62.5bp)=(0.69,0.81,0.97);
			rgb(100bp)=(0.69,0.81,0.97)}
		
		% Gradient Info
		
		\tikzset {_lk4l2ho6q/.code = {\pgfsetadditionalshadetransform{ \pgftransformshift{\pgfpoint{0 bp } { 0 bp }  }  \pgftransformrotate{0 }  \pgftransformscale{2 }  }}}
		\pgfdeclarehorizontalshading{_i7xgx01sd}{150bp}{rgb(0bp)=(0.96,0.65,0.14);
			rgb(37.5bp)=(0.96,0.65,0.14);
			rgb(62.5bp)=(0.69,0.81,0.97);
			rgb(100bp)=(0.69,0.81,0.97)}
		
		% Gradient Info
		
		\tikzset {_hmouzgu57/.code = {\pgfsetadditionalshadetransform{ \pgftransformshift{\pgfpoint{0 bp } { 0 bp }  }  \pgftransformrotate{0 }  \pgftransformscale{2 }  }}}
		\pgfdeclarehorizontalshading{_5uf4s424d}{150bp}{rgb(0bp)=(0.69,0.81,0.97);
			rgb(37.589285714285715bp)=(0.69,0.81,0.97);
			rgb(62.5bp)=(0.94,0.94,0.96);
			rgb(100bp)=(0.94,0.94,0.96)}
		
		% Gradient Info
		
		\tikzset {_5zbegqp8n/.code = {\pgfsetadditionalshadetransform{ \pgftransformshift{\pgfpoint{0 bp } { 0 bp }  }  \pgftransformrotate{0 }  \pgftransformscale{2 }  }}}
		\pgfdeclarehorizontalshading{_hojaglvxc}{150bp}{rgb(0bp)=(0.69,0.81,0.97);
			rgb(37.589285714285715bp)=(0.69,0.81,0.97);
			rgb(62.5bp)=(0.94,0.94,0.96);
			rgb(100bp)=(0.94,0.94,0.96)}
		
		% Gradient Info
		
		\tikzset {_x69a3p5dz/.code = {\pgfsetadditionalshadetransform{ \pgftransformshift{\pgfpoint{0 bp } { 0 bp }  }  \pgftransformrotate{0 }  \pgftransformscale{2 }  }}}
		\pgfdeclarehorizontalshading{_3r9gk39o2}{150bp}{rgb(0bp)=(0.82,0.7,0.93);
			rgb(37.589285714285715bp)=(0.82,0.7,0.93);
			rgb(62.5bp)=(0.94,0.94,0.96);
			rgb(100bp)=(0.94,0.94,0.96)}
		
		% Gradient Info
		
		\tikzset {_tysm95dji/.code = {\pgfsetadditionalshadetransform{ \pgftransformshift{\pgfpoint{0 bp } { 0 bp }  }  \pgftransformrotate{0 }  \pgftransformscale{2 }  }}}
		\pgfdeclarehorizontalshading{_t8b4p12cq}{150bp}{rgb(0bp)=(0.82,0.7,0.93);
			rgb(37.589285714285715bp)=(0.82,0.7,0.93);
			rgb(62.5bp)=(0.94,0.94,0.96);
			rgb(100bp)=(0.94,0.94,0.96)}
		
		% Gradient Info
		
		\tikzset {_xzrzrpnm5/.code = {\pgfsetadditionalshadetransform{ \pgftransformshift{\pgfpoint{0 bp } { 0 bp }  }  \pgftransformrotate{0 }  \pgftransformscale{2 }  }}}
		\pgfdeclarehorizontalshading{_0kyqig4xw}{150bp}{rgb(0bp)=(0.82,0.7,0.93);
			rgb(37.589285714285715bp)=(0.82,0.7,0.93);
			rgb(62.5bp)=(0.7,0.8,0.93);
			rgb(100bp)=(0.7,0.8,0.93)}
		
		% Gradient Info
		
		\tikzset {_gx5ts9y5q/.code = {\pgfsetadditionalshadetransform{ \pgftransformshift{\pgfpoint{0 bp } { 0 bp }  }  \pgftransformrotate{-90 }  \pgftransformscale{2 }  }}}
		\pgfdeclarehorizontalshading{_ltrn0afbe}{150bp}{rgb(0bp)=(0.92,0.95,0.79);
			rgb(37.5bp)=(0.92,0.95,0.79);
			rgb(62.5bp)=(0.82,0.7,0.93);
			rgb(100bp)=(0.82,0.7,0.93)}
		\tikzset{every picture/.style={line width=0.75pt}} %set default line width to 0.75pt        
		
}
	%\vspace{-0.25cm}
	\caption{The hybrid metaheuristic-ML framework for UE-AP association with joint resource allocation and blockage~avoidance.}
	\label{Fig:Hybrid_AI_ML}
\end{figure*}

As illustrated in Fig.~\ref{Fig:Hybrid_AI_ML}, the approach has an offline learning phase, which is used for training and validating the ML model responsible for UE-AP association and RR allocation, and an online learning phase that ensures adaptation to the dynamic wireless network. In the offline learning phase, a dataset that contains different instances of the network topology, i.e., UE, AP, and blockers positions, each AP RRs and each UE data-rate demands, is input to a metaheuristic optimizer. The optimizer employs a number of state-of-the-art algorithms, like tabu search (TS), simulated annealing (SA), late acceptance (LA), and strategic oscillation (SO) to extract the optimal UE-AP association, where the RR allocation and blockage avoidance constraints are satisfied for each~instance.

The metaheuristic optimizer's output is split into training and validation data. The training data are forwarded into a relative labeling (RLB) scheme, which categorizes all APs in the network. The categorization creates three sub-classes that represent essential aspects of APs. These sub-classes are concatenated together to create a singular class, resulting in a multi-class label attribute. These sub-classes are created as~follows:
\begin{itemize}
 \item\textit{Geographic Location}: Using APs' coordinates, we divide each AP among 4 quadrants, namely: North-East, North-West, South-East, and~South-West.
 \item \textit{Resource Capacity}: Considering the available RR among all APs, we create 10 sub-classes that quantize the RR capacity of each~AP.
 \item \textit{LoS Blockages}: Considering the number of partial blockages that the $i-$th UE would face when establishing a direct connection with the $j-$th AP, we create 9 sub-classes that quantize the level of partial~blockage.
\end{itemize}

Next, by treating the AP that was associated by metaheuristic optimization to a UE as ground truth, we append the corresponding RLB of this AP to complete the training dataset. The ML model is trained to predict this RLB, which are mapped back to an AP in a post-prediction step. Several ML models including generalized linear models (GLM), naive Bayes (NB), random forests (RF), gradient boosted trees (GBT), and deep learning (DL) can be employed. The model is validated using the validation data produced by the metaheuristic~optimizer. 

In the online learning phase, the trained ML model is used to perform UE-AP association and RR allocation on unseen snapshots and the predicted solutions are compared to the ground truth. In case of unacceptable performance the model is retrained on the new snapshots.   

\underline{Indicative results}: We use the following scenario To demonstrate the performance of our approach and highlight some of the design decisions involved. A wireless network that operates in the $100\,\mathrm{GHz}$ band and consists of $741$ UEs, $125$ APs in a $1\,\mathrm{km}^2$ disk-shaped area. The positions of UEs and APs are determined by independent Poisson point processes. The available RR at each AP, as well as the UE requested data-rates are modeled as independent uniform random variables. Both the APs and UEs locations are described by polar coordinates. A third element of the model is a tuple that describes an initial minimum-distance-basted UE-AP clustering. The cluster is obtained using k-means without accounting for the LoS blockage. This clustering reveals UEs that potentially block each others' LoS when facing a particular~AP. 

In the simulated scenario the theoretic upper bound of possible AP-UE combinations is $10^{1553}$. This quantifies the association complexity. The average blockage density gives the number of blockers a single UE would face on average when a direct link with an AP is considered. Finally, the pairs of  UE-AP that have at least 1 UE blocking their LoS is $62.5\%$. Hence, the dataset presents a formidable challenge for the metaheuristic algorithm(s) to optimize the specified~constraints. 

\begin{figure}
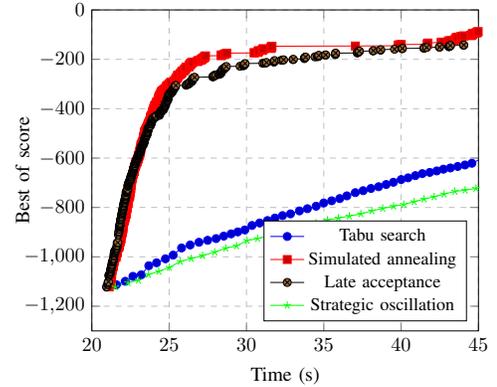
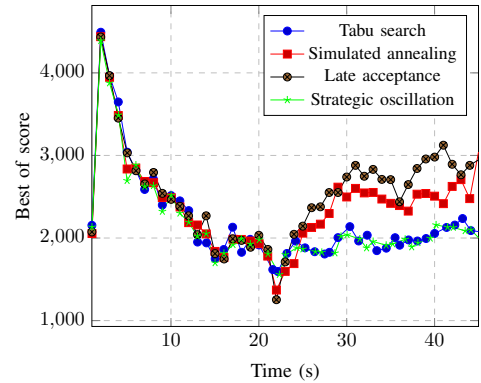

	\centering
	\subfloat[\centering Progress of optimization algorithms]{{{\scalebox{0.75}{
}} }}%
	\vspace{+0.5cm}
	\subfloat[\centering Solutions processed per second ]{{\scalebox{0.75}{%
} }}%
	\caption{Metaheursitic optimizations to discover UE-AP associations}%
	\label{Fig:AIpaper-MetaheuristicOptimizations}%
\end{figure}

Figure~\ref{Fig:AIpaper-MetaheuristicOptimizations} quantifies the performance of the different types of metaheuristic optimizers. Figure~\ref{Fig:AIpaper-MetaheuristicOptimizations}.a illustrates how the search algorithms' improvement over time (w.r.t. the optimization score), while Fig.~\ref{Fig:AIpaper-MetaheuristicOptimizations}.b depicts the traversal of the search space in terms of solutions processed per second. These results provide insights on how different algorithms perform given a specific dataset and optimization constraints. The two figures reveal that SA and LA dominate TS and~SO.

After labeling the optimizer's outputs, we perform a 95-5 split, i.e., $95\%$ data was split (using stratified sampling) for training and $5\%$ kept for validation and we test the performance of GLM, NB, RF, GBT, and DL. Overall, the results of GBT are superior. The accuracy of the model reaches $99.41\%$ with precision and recall for all classes above $98\%$ and standard deviation of the model in the order of $0.15\%$, which testifies to the stability of the model's performance. The accuracy upon validating the model is $99.29\%$, which shows that the model achieve acceptable performance under unseen but similar network~snapshot.

\begin{table*}
	\centering
	\caption{UE-AP associations as discovered by the LA algorithm and predicted by the GBT model.}
	\label{Table:AI-Paper-solutions-table}
	\resizebox{\textwidth}{!}{
		\begin{tabular}{|l|lll|llll|}
			\hline
			\multicolumn{1}{|c|}{\textbf{Method}} & \multicolumn{3}{c|}{\textbf{UEs (total=741)}}                                                                                   & \multicolumn{4}{c|}{\textbf{APs (total=125)}}                                                                                                                           \\ \hline
			& \multicolumn{1}{l|}{\textbf{Allocated}} & \multicolumn{1}{l|}{\textbf{Unblocked Links}} & \textbf{Links with 1 Partial Blocker} & \multicolumn{1}{l|}{\textbf{Used}} & \multicolumn{1}{l|}{\textbf{Capacity Respected}} & \multicolumn{1}{l|}{\textbf{Capacity Overloaded}} & \textbf{Average UEs per AP} \\ \hline
			Metaheuristics (LA)                   & \multicolumn{1}{l|}{741}                & \multicolumn{1}{l|}{741}                      & 0                                     & \multicolumn{1}{l|}{95}            & \multicolumn{1}{l|}{70}                          & \multicolumn{1}{l|}{25}                           & 7.8                         \\ \hline
			ML (GBT)                      & \multicolumn{1}{l|}{741}                & \multicolumn{1}{l|}{643}                      & 98                                    & \multicolumn{1}{l|}{98}            & \multicolumn{1}{l|}{78}                          & \multicolumn{1}{l|}{20}                           & 7.56                        \\ \hline
		\end{tabular}
	}
\end{table*}

To evaluate the solution's QoE  for both metaheuristic optimization and ML side, Table~\ref{Table:AI-Paper-solutions-table}, given at the top of the next page, summarizes functional and non-functional key performance indicators (KPIs). From this table, it becomes evident that both metaheuristic and ML were able to assign APs to all 741 UEs. This shows that both methods have a good capability to provide network coverage. Moreover, the balance of both approaches is quantified. Note that ``balance'' shows how the resource use was balanced by using a certain number of APs. {If lesser than total APs are used effectively, then energy consumption can also be reduced, which reduces the carbon footprint.} We observe that the metaheuristic solution is more efficient as it employs $95$ APs, while the ML solution uses $98$ APs to satisfy the  resources required by all the~UEs.

The QoE is determined by the provisioning of required RR and avoidance of LoS blockages. RR utilization can measured by overload or underload of capacity among the used APs. We observe that the metaheuristic solution has slightly overloaded $25$ APs with overall capacity overloaded by $4.4\%$ on average. The ML solution has overloaded only $20$ APs, but the overall capacity has overloaded by $14.5\%$ on average. This shows that although the ML model exhibits that it has recognized and learned to respect the resource satisfaction constraint, yet further training is needed. Moreover, the QoE is also influenced by the number of LoS blockers that are avoided or minimized. The metaheuristic optimizer was able to discover APs for all $741$ UEs such that no partial blockers existed in the direct LoS link between them. In case of the ML, a vast majority of $643$ UEs were associated with APs with no LoS blocker and $98$ UEs were associated with APs, where the LoS blockers were minimized to $1$. This result further confirms that the model learnt the pattern to avoid LoS blockages or minimizing~them.

Latency is a non-functional criterion, which was not intended to evaluate the optimization performance, because the purpose of the offline optimization is merely to discover optimal UE-AP associations for the generation of training data. The ML model trained using this data is the actual artifact. Hence, the inference efficiency of the ML model for a given data-point is of actual interest to conclude whether the overall connection establishment latency meets the B5G level requirement. A data-point is a vector of UE features and the network snapshot at a particular point in time. The scoring (or inference) time is dependent on the hardware, where the ML model is deployed. In our case, the scoring time was $0.75\,\mathrm{ms}$ for $1$ data-point, which shows a promising connection latency. However, if the GBT model is further pruned and feature engineering can further reduce the dimensions of data, then this time can be further reduced.
We also recognize that it would be better to obtain predictions for multiple data-points, which means that a total of data-points that is equal to the number of UEs multiplied by the number of APs, needs to be predicted. The relative label predicted most often can then be reliably used to identify the optimal AP. In our case, for the vast majority of UEs, the same relative label was predicted even with fewer data-points. When scoring against all APs, predictions took $94.5\,\mathrm{ms}$ on average. However, domain-specific knowledge can be applied to limit the number of APs e.g., to the nearest 3 APs, in which case, the AP association can be predicted in~$2.26\,\mathrm{ms}$.

\section{DRL~for Adaptive Beam~Allocation}

\underline{Problem description}: Within each AP serving area we consider a dynamic wireless network, where UEs and blockers are allowed to changed positions. An AP needs to be equipped with an intelligent mechanism to solve RR management/allocation problems. This motivates the idea of having the AP interact with the environment by using deep reinforcement learning (DRL). In particular, DRL can be used as a tool for sum-rate maximization through beamforming optimization in multi-UE THz wireless~networks.

\underline{A deep reinforcement learning approach}:
Both DRL and reinforcement learning (RL) are methods for creating software agents that can learn to interact with environments in an unsupervised manner. Instead of learning from a labeled dataset, DRL receives feedback in the form of a reward. By changing its behavior to maximize the cumulative reward over multiple steps, the agent improves its performance~\cite{Sutton2017}. At each step, the agent receives a representation of the environment’s state and it predicts an optimal action to take given this state. The action is in turn used to update the state and to compute a reward value. This process continues until the agent reaches some signified end state or until some other stopping criterion (e.g. a maximum number of iterations) is met~\cite{Sutton2017}.

\begin{figure}
	\centering
	\subfloat[\centering]{{{\includegraphics[width=1.0\linewidth,trim=0 0 0 0,clip=false]{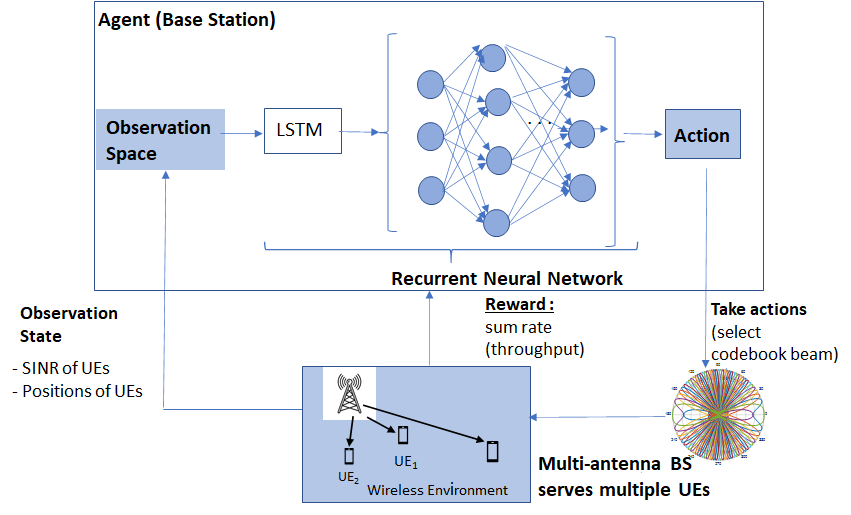}} }}%
	\vspace{+0.5cm}
	\subfloat[\centering]{{{\includegraphics[width=1.0\linewidth,trim=0 0 0 0,clip=false]{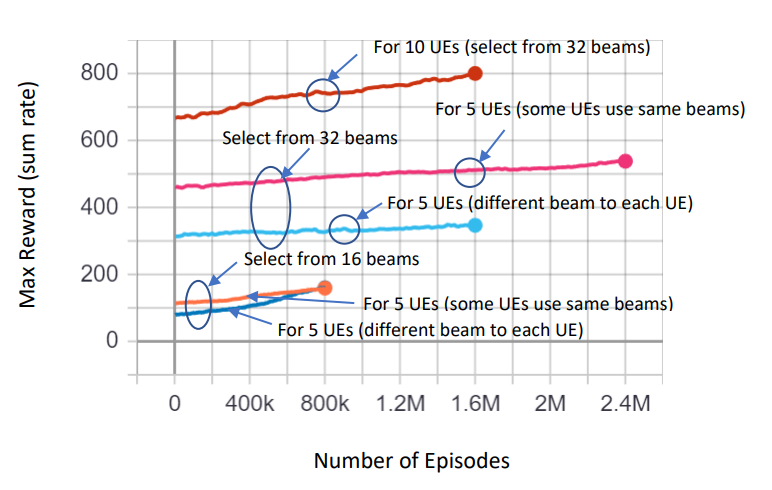}} }}%
%	\vspace{+0.5cm}
	\caption{a) Information flow between agent and environment with action space and observation space, and b) Reward versus number of episodes.}
	\label{fig:DRL_information_flow}
\end{figure}

In our case, an agent corresponds to the AP and the actions correspond to codebook selection. By learning how to take actions, the AP can select the codebook beam from the predefined codebook beamset, while serving the UEs at the same time. The objective is to maximize the cumulative reward, which is defined as the sum of the UEs data rates. The environment (observation space) is modelled by information related to the signal-to-interference-plus-noise-ratio (SINR) of the UEs and the UEs' positions over time. Environmental states generated over time by the agent's interaction with the environment, as outlined above, are input to an LSTM network (this particular type of network is capable of encoding in its weights long-range temporal relations between states), which learns to sequentially  predict actions guided by the sequence of received rewards over a training session. The overall approach and information flow is illustrated in Fig.~\ref{fig:DRL_information_flow}.a. 

\underline{Implementation based on Proximal Policy Gradient (PPO)}:
We adopt the state-of-the-art PPO framework \cite{schulman2017proximal} for implement our DRL approach. In policy gradient methods, a neural network is used to represent a policy function, which maps an environment state directly into an action. The network parameters are then updated during training to maximize the expected cumulative reward produced by the actions. When computing the gradients for policy updates, the value function can be used together with the sampled rewards to improve the quality of the updates. The combination of policy and value functions into one RL agent is called the ``actor-critic'' model. While training the network, the actor performs actions according to its current policy, which is used to select the actions and the estimated value function. After each action selection, the critic provides feedback based on how good or bad it considers those actions to~be.

PPO uses the actor-critic model to learn the policy. The PPO architecture enables the learner to be accelerated using graphic processing units and actors to be distributed across several machines~\cite{Schulman_PPO}. Then, we use RLiib, which is a module in ray distribution package with implementations of RL~algorithms.

\underline{Results}: Figure~\ref{fig:DRL_information_flow}.b illustrates the maximum reward versus number of episodes of the system. We consider two main scenarios which are (i) when the agent (AP) selects different codebook beam to serve each UE and (ii) when the AP allows more than one UE to use the same~beam. In this system, we consider that the AP allocates codebook beam to serve $5$ UEs and $10$ UEs at the same time. We assume that the AP deploys $8$ antennas with $16$ predefined beams, and in the other set-up, the AP deploys $16$ antennas with $32$ predefined beams. The UEs are randomly located near by the AP, i.e., within $100$ metres and the UEs are near to each other. We can see that when some UEs can be served by the same beam, the reward is higher than when the AP allocates different beam to each UE. The reason is that some UEs locate close to each other want to use the same beam; therefore, this leads to higher~reward.

\section{AI-assisted  fast-channel prediction for pro-active hand-over}

\underline{Problem description}: In an ultra-dense wireless network, where several APs are capable of serving the same MUE, in order to maximize the system's reliability, efficiency, and throughput, it is important to enable pro-active hand-over based on accurate channel characteristics predictions. The following features have been identified as key inputs of most hand-over policies in the THz band: i) signal strength, ii) directionality, and iii) existence of LoS path. Conventional approaches that have been extensively employed in lower frequency bands are rendered infeasible due to inherent complexity that is originated from the high number of degrees of freedom of massive multiple-input multiple-output (MIMO)~systems. 

\underline{ML-based approach}: To overcome this barrier, ML-based models that extract the relationship between the environment and the aforementioned channel characteristics, are expected to become key enablers of URLL proactive hand-over approaches. The first step towards creating such an ML model is to acquire training and testing data. In this direction, we collected the baseband channel coefficients, delays, {azimuth angle of departure (AAoD), elevation angle of departure(EAoD),} {azimumth angle of arrival (AAoA), elevation angle of arrival (EAoA),} delay spread angle (DSA), and azimuth angle (AA) as well as the transmitter and receiver characteristics of a  MUE that performed a number of different routes. These data were fed in a pre-processing unit that returned their statistical characteristics, namely {expectations and~extension}.  

Since the  available dataset has a limited feature set and observed error margins, the approach relies on investigating the existence of patterns within the available features, as well as taking into account the geographic location, in order to examine the density according to the user mobility for predicting LOS connectivity. Initial results reveal a certain degree of correlation among attributes, when location information is included, but the predictive modeling aspects require additional feature engineering towards achieving various KPIs related to accuracy, fading statistics, correlations, complexity and versatility, which can find the hidden non-linear relationships among different features and can indeed improve the feasibility and viability of the~model.

This inspires breaking the channel prediction problem into two sub-problems. The first one focuses on the behavior of raw or ﬁne granular form of individual path data, while the second one aims to aggregate (coarse granular) data, subject to feature engineering. The former can be used to predict the presence of the LOS or non-LOS path, path-loss for higher gain, or higher power, delay spread, etc.  The latter is capable of predicting the link level behaviors, e.g. the properties of a link proportional to the received~power for a given location. 

Both sub-problems follow similar solution approaches that include pre-processing, simulation execution, and post-processing. 
Pre-processing  involves model settings and input data  preparation for further training so that the data can be used easily and effectively. This steps often ends when a simulation is ready to launch. The simulation phase will allow interactive visualization to be performed at each time step. In more detail, based on user mobility across a certain route in specific environment, it will show whether the route on which the user is moving falls under LOS or non-LOS connectivity. Moreover, the simulations tailor down how the model performs with respect to reality. Post-processing extracts the results and put them in usable~form.

 The dataset was split into training and testing sets. We performed Feature selection and parameter tuning and trained various ML algorithms to assess which types of model achieve superior performance. In general, ML helps to extract significant features from the  data to deliver insightful predictions. Various ML models are built and validated on testing data . The ML, such as decision tree, RF, NB for classifying LoS vs NLoS identification, are compared to evaluate the~performance. 
 
 \underline{Results}: Decision tree achieves the best performance in terms of precision. Table~\ref{Table:Confusion_matrix_results} depicts the prediction performance in form of confusion matrix, which shows the models compares resulting classification outcome with actual values of the given observation. There are four potential outcomes here: true positive (TP) value 3072 indicates the model predicted an outcome of true, and the actual observation was true. False positive (FP) value $55$ indicates the model predicted a true outcome, but the actual observation was false. False negative (FN) value $5673$ indicates the model predicted a false outcome, while the actual observation was true. Finally, we have the true negative (TN) value $14076$, which indicates the model predicted an outcome of false, while the actual outcome is also false. Accuracy is calculated, which is generally used to judge the model performance. However, in this particular  balanced dataset, the accuracy of $75\%$ was achieved with decision~tree. 

\begin{table}
	\centering
	\caption{Confusion matrix for performance assesment.}
	\label{Table:Confusion_matrix_results}
	%\resizebox{\textwidth}{!}{
		\begin{tabular}{|l|l|l|l|}
			\hline
			& {\textbf{True LoS}} & {\textbf {True NLoS}} & {\textbf{Precision}} \\ \hline
			{\textbf{Pred. LoS}} & 3072 & 55 & 90.24\%\\ \hline
			
			{\textbf{Pred. NLoS}} & 5673 & 14076 & 70.85\%\\ \hline
			{\textbf{Recall}} & 34.71\% & 99.61\% &   
			
			\\ \hline
			
		\end{tabular}
		%}
\end{table}

The results shows room for improvements, which can be explicitly made by using optimizing parameters to further fine tune the parameters in order to extract and learn more patterns for the model. The complexity of these tasks is frequently beyond the capabilities of non-domain experts; hence, the quick development of ML applications has increased the demand for ready-to-use ML techniques. It is the ensuing field of study that focuses on progressive automation of ML, which compares multiple models, set the default parameters all at once and outputs the accuracy, run time error for different~algorithms. 
\begin{figure}
\centering
	\includegraphics[width=0.9\linewidth]{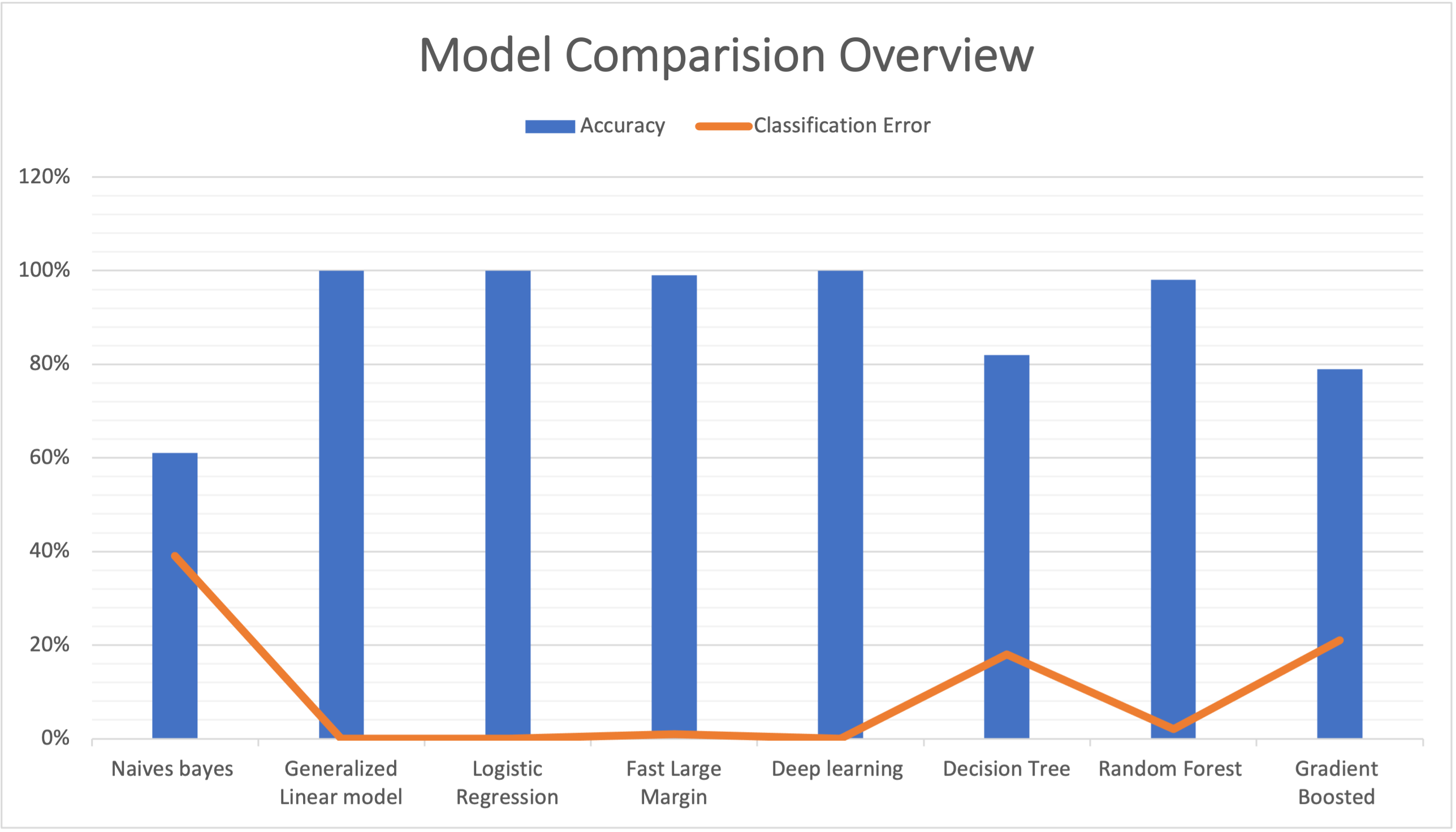}
	\caption{Model comparison.}
	\label{Fig:AutoModel Results Evaluation}
\end{figure}

{Figure~\ref{Fig:AutoModel Results Evaluation} illustrates the performance of different models in terms of accuracy, along with its respected classification run time error for each model. As the execution time error decreases, the model improves. 
Hence, uncovering the inﬂuence of environmental characteristics and other micro-and macro-cellular scenarios that are essential for evaluating the potential of two polarized MIMO systems and the efficacy of LOS connectivity will be the goals worth pursuing to see if they outperform the conventional~methods.}

\section{Conclusions}

In this article, we presented a holistic and intelligent MAC layer approach that enhances THz wireless systems reliability, while minimizes the association latency, through a metaheuristic-ML based joint UE-AP association and RR allocation with blockage avoidance mechanism, support UE mobility and blockage avoidance within the AP coverage area through a DRL beam selection scheme, and enables ultra-reliable and pro-active hand-over via AI-assisted fast-channel prediction. The results provided useful guidelines for the design of intelligent MAC for URLL THz wireless systems and highlighted the key role of ML in these~networks.     

\balance
\bibliographystyle{IEEEtran}
\bibliography{IEEEabrv,References}

\begin{IEEEbiography}{Alexandros-Apostolos A. Boulogeorgos}  received the Electrical and Computer Engineering diploma and Ph.D. degrees both from the Aristotle University of Thessaloniki in 2012 and 2016, respectively. Currently, he works as a postdoctoral researcher in the University of Piraeus. He has published more than $100$ contributions in journals and conferences. His research interests include wireless and optical wireless communication theory, reconfigurable intelligent surface-assisted communications, and wireless communications for biomedical applications. He serves as an Editor in IEEE COMMUNICATIONS LETTERS, Frontiers in Communications and Networks, and MDPI Telecom. 
\end{IEEEbiography}

\begin{IEEEbiography}{Edwin Yaqub} is a Lead Data Scientist working with RapidMiner since 2016. He has expertise in artificial intelligence, machine learning, constraint solving optimizations, data enrichment and product prototyping, which he has successfully applied to solve a variety of industrial and scientific problems. He amassed 13 years of experience in research and commercial projects from diverse sectors including finance, automotive, healthcare, eCommerce and telecommunication.
Edwin received his MSc. degree in Computer Science in 2008 from RWTH Aachen University, Germany. His Master thesis at Philips Research Labs developed a framework for automated health decision support using ontologies and rules-based reasoning. In 2015, he earned Ph.D. in Computer Science from the University of Göttingen, Germany. In his doctoral research, he investigated decision-making algorithms to automatically negotiate and optimize provisioning of cloud infrastructure resources. His recent work addresses resource management using optimizations, predictive and prescriptive machine learning in B5G and 6G wireless networks.
\end{IEEEbiography}

\begin{IEEEbiography}{Rachana Desai}, Senior Data scientist at RapidMiner  with extensive experience from the field of Bioinformatics. She completed  her M.Sc. in General Engineering with a focus on bioinformatics at the San Jose University(California, USA) with  International publications demonstrating expertise in the field of data science as part of agile development teams. During her Master’s degree she worked on optimization and dimension reduction of  3D-QSAR validation metrics and linear SVM cross-validation metrics to facilitate by comparing the best pre-prediction bio-statistical models where she developed interest in machine learning and AI. Since 2019, she is working with RapidMiner on data science projects covering vivid industries  like electronics and telecommunication.  She worked previously with Vizzario, Inc. and Cisco Systems,Inc. as a data scientist  in USA with expertise in Implementing, designing and developing algorithms on Machine Learning, Optimization Algorithms, Deep learning and AI by solving customer problems in areas such as wireless network troubleshooting, physical surveillance, security, IoT visual sensing and retail analytics.
\end{IEEEbiography}

\begin{IEEEbiography}{Tachporn Sanguanpuak} received B.S. degree in telecommunications engineering from King Mongkut's Institute of Technology Ladkrabang (KMITL), Thailand in 2010, M.S. degree in telecommunications engineering from Asian Institute of Technology (AIT), Thailand in 2012. She received Ph.D. degree in telecommunications engineering from Centre for Wireless Communications, University of Oulu, Finland in 2019. She joined Nokia RF/MN, Oulu, Finland in 2020 and then joined Nokia standardization team, Oulu, Finland in June 2022. She held senior specialist radio research position in standardization team, and she mainly works for 3GPP standardization. Her research interests include radio resource management and machine learning for wireless communications.
\end{IEEEbiography}

\begin{IEEEbiography}{Nikos Katzouris} is a research associate at the Institute of Informatics \& Telecommunications, National Center for Scientific Research ``Demokritos'', in Athens, Greece. His research is mostly focused on machine learning and artificial intelligence, and in particular, on combinations of machine learning with knowledge representation and reasoning and applications of such hybrid techniques to the field of complex event recognition and forecasting. He holds an PhD in machine learning, an Msc in logic \& computation and a Bsc in mathematics.  
\end{IEEEbiography}

\begin{IEEEbiography}{Fotis Lazarakis} received his diploma in Physics, from Department of Physics, Aristotle University of Thessaloniki, Greece (1990), and his Ph.D in Mobile Communications from Department of Physics, National and Kapodistrian University of Athens, Greece (1997), holding at the same time a scholarship from National Center for Scientific Research ``Demokritos'' (NCSRD), Institute of Informatics and Telecommunications (IIT). From 1999 to 2002 he was with Telecommunications Laboratory, National Technical University of Athens, as a senior research associate. In 2003 he joined NCSRD, IIT as a Researcher and since 2013 is a Research Director. He has been involved in a number of national and international projects, acting as a Project Manager to several of those. His current research interests include 5G and beyond, as well as 6G radio technologies and Reconfigurable Intelligent Surfaces. Dr. Lazarakis has authored or co-authored 120 journal and conference papers and he is co-owner of a patent.
\end{IEEEbiography}

\begin{IEEEbiography}{Angeliki Alexiou} is professor of Broadband Communications Systems at the department of Digital Systems, University of Piraeus, Greece. She received the Diploma in Electrical and Computer Engineering from the National Technical University of Athens (1994) and the PhD in Electrical Engineering from Imperial College, University of London (2000). Since May 2009 she has been a faculty member at the Department of Digital Systems. Prior to this appointment (January 1999-April 2009) she was with Bell Laboratories, Wireless Research, Lucent Technologies, (now NOKIA), UK. Professor Alexiou is a co-recipient of Bell Labs President’s Gold Award  (2002) for contributions to Bell Labs Layered Space-Time (BLAST) project and the Central Bell Labs Teamwork Award (2004) for role model teamwork in the IST FITNESS project. Professor Alexiou is the Chair of the Working Group on Radio Communication Technologies of WWRF. Her current research interests include advanced radio for 6G, THz communications and Reconfigurable Intelligent Surfaces. 
\end{IEEEbiography}

\begin{IEEEbiography}{Marco Di Renzo} is a CNRS Research Director (Professor) with the
Laboratory of Signals and Systems of Paris-Saclay University – CNRS
and CentraleSupelec, Paris, France. He serves as the Coordinator of
the Communications and Networks Research Area of the Laboratory of
Excellence DigiCosme a and as a Member of the Admission and Evaluation
Committee of the Ph.D. School on Information and Communication
Technologies at Paris-Saclay University, and as the Head of the
Intelligent Physical Communications group with the Laboratory of
Signals and Systems at CentraleSupelec. He is the Editor-in-Chief of
IEEE Communications Letters. He is a Fellow of the IEEE, IET, and
AAIA; an Ordinary Member of the European Academy of Sciences and Arts,
and the Academia Europaea; and a Highly Cited Researcher. Also, he is
a Fulbright Fellow. Recent research awards include the 2022 IEEE
COMSOC Outstanding Paper Award and the 2022 Michel Monpetit Prize from
the French Academy of Sciences.
\end{IEEEbiography}

\end{document}